\documentclass[graybox]{svmult}

\usepackage{type1cm}        
\usepackage{makeidx}         
\usepackage{graphicx}        
\usepackage{multicol}        
\usepackage[bottom]{footmisc}

\usepackage{newtxtext}       %
\usepackage{newtxmath}       

\usepackage{mathtools}
\usepackage{physics}
\usepackage{bm}
\usepackage{booktabs}
\usepackage{multirow}
\usepackage{url}
\usepackage[position=top, caption=false]{subfig}
\usepackage{tikz}
\usetikzlibrary{quantikz2}
\usepackage{hyperref}
\usepackage{cleveref}
\usepackage[numbers,sort&compress]{natbib}

\makeatletter
\newcommand*\bigcdot{\mathpalette\bigcdot@{.5}}
\newcommand*\bigcdot@[2]{\mathbin{\vcenter{\hbox{\scalebox{#2}{$\m@th#1\bullet$}}}}}
\makeatother

\makeatletter
\DeclareFontFamily{OMX}{MnSymbolE}{}
\DeclareSymbolFont{MnLargeSymbols}{OMX}{MnSymbolE}{m}{n}
\SetSymbolFont{MnLargeSymbols}{bold}{OMX}{MnSymbolE}{b}{n}
\DeclareFontShape{OMX}{MnSymbolE}{m}{n}{
    <-6>  MnSymbolE5
   <6-7>  MnSymbolE6
   <7-8>  MnSymbolE7
   <8-9>  MnSymbolE8
   <9-10> MnSymbolE9
  <10-12> MnSymbolE10
  <12->   MnSymbolE12
}{}
\DeclareFontShape{OMX}{MnSymbolE}{b}{n}{
    <-6>  MnSymbolE-Bold5
   <6-7>  MnSymbolE-Bold6
   <7-8>  MnSymbolE-Bold7
   <8-9>  MnSymbolE-Bold8
   <9-10> MnSymbolE-Bold9
  <10-12> MnSymbolE-Bold10
  <12->   MnSymbolE-Bold12
}{}

\let\llangle\@undefined
\let\rrangle\@undefined
\DeclareMathDelimiter{\llangle}{\mathopen}%
                     {MnLargeSymbols}{'164}{MnLargeSymbols}{'164}
\DeclareMathDelimiter{\rrangle}{\mathclose}%
                     {MnLargeSymbols}{'171}{MnLargeSymbols}{'171}
\makeatother

\makeindex             

\begin{document}

\title*{Quantum Error Correction and Detection for Quantum Machine Learning}
\author{Eromanga Adermann, Haiyue Kang, Martin Sevior and Muhammad Usman}
\institute{Eromanga Adermann \at Quantum Systems, CSIRO, Marsfield, NSW, 2122, Australia 
\and
Haiyue Kang \at School of Physics, University of Melbourne, VIC, Parkville, 3010, Australia
\and
Martin Sevior \at School of Physics, University of Melbourne, VIC, Parkville, 3010, Australia
\and
Muhammad Usman \at Quantum Systems, CSIRO, Clayton, VIC, 3178, Australia \at School of Physics, University of Melbourne, VIC, Parkville, 3010, Australia  
}

%
%
\maketitle

\abstract{
At the intersection of quantum computing and machine learning, quantum machine learning (QML) is poised to revolutionize artificial intelligence. However, the vulnerability of the current generation of quantum computers to noise and computational error poses a significant barrier to this vision. Whilst quantum error correction (QEC) offers a promising solution for almost any type of hardware noise, its application requires millions of qubits to encode even a simple logical algorithm, rendering it impractical in the near term. In this chapter, we examine strategies for integrating QEC and quantum error detection (QED) into QML under realistic resource constraints. We first quantify the resource demands of fully error-corrected QML and propose a partial QEC approach that reduces overhead while enabling error correction. We then demonstrate the application of a simple QED method, evaluating its impact on QML performance and highlighting challenges we have yet to overcome before we achieve fully fault-tolerant QML.}

\section{Introduction}
Quantum machine learning (QML)\cite{QML,QML2,QML3, VQA,VQE,QAOA,QPCA,QNN, wang2024quantum, tsang2023hybrid, wang2025self, gill2022quantum, khatun2025classical, transfer} is an emerging field that promises performance improvements over classical counterparts, including training speed-ups, increased reliability and superior feature extraction capabilities. In recent years, a variety of QML models \cite{QML_full_summary,QML,QML2,VQA,VQE,Accelerated_VQE,VQE_variants_summary,VQSE,QAOA,QPCA,QNN,QNN_2,VQC,QVC,QVC_2,QGAN,QGAN2,TN_QML,quantum_chemistry,QCNN_classically_simulable,reflection_invariant_qml,rotational_invariant_qml,non_unitary_qml,Zehang_Wang_QML_1,barren_plateaus,Max_West_adv_QML_1,Max_West_adv_QML_2,emprical_VQE,thermal_helps_learning,thermal_helps_learning2} have been developed and characterized for various datasets, with the most widely adopted model being the quantum variational classifier (QVC) \cite{QVC}. The basic working principle of QVC-based QML is quite similar to the classical machine learning in the sense that a loss function is iteratively minimized, driving towards its global minimum, which is usually achieved in the form of fitting some highly non-trivial parametrized function. 

However, the prevalence of noise in the current generation of quantum processors poses a significant challenge for practical-scale implementations of QML, as they lead to trainability issues arising from noise-induced barren plateaus \cite{noise_induced_barren_plateaus} and performance degradations from noise accumulation in deep circuits. Quantum error correction (QEC) protocols are under development as a solution to hardware noise, but their extremely high spacetime overheads \cite{surface_codes,qec_lattice_surgery}, due primarily to magic state distillation, make them infeasible for near-term practical implementation. 

As such, we are presented with the challenge of mitigating, detecting and correcting errors that are bound to arise with implementations of QML on quantum hardware with physical resource limitations. To this end, we investigate practical strategies for integrating QEC into QML. We begin by detailing the resources required to realise full quantum error corrected quantum machine learning, to illustrate the extent of the challenge before us. We then explore a means of circumventing this challenge with the partial application of QEC, which allows for a significant reduction in spacetime overhead. We conclude this chapter by demonstrating the application of a simple error detection code to a QML algorithm and evaluating its effectiveness. 

\section{Resource overhead for quantum error correction}
The implementation of quantum machine learning (QML) with quantum error correction (QEC) can address challenges associated with trainability and noise accumulation in deep quantum circuits. However, existing QEC protocols, such as surface codes, are expensive in terms of spacetime cost for QML architectures. Variational QML circuits rely on parametrized quantum gates, which are non-Clifford operations and typically require highly expensive logical $T$-gate implementation through magic state injection, which is based on the preparation of a logical magic state from a single physical qubit or short-distance qubit first \cite{surface_codes}. Without the redundancy of multiple physical qubits, the logical $T$-gate error rate is close to the physical qubit error rate. Thus, one must mitigate the magic state through state distillation \cite{surface_codes, magic_state_distillation_cost1, magic_state_distillation_cost2}, yet this leads to an exponentially large spacetime overhead. To quantitatively address how large the cost is, we present an analysis of the spacetime overhead of executing a full 10-qubit quantum variational circuit with QEC.

\begin{figure}[h]
    \centering
    \includegraphics[width=.8\linewidth]{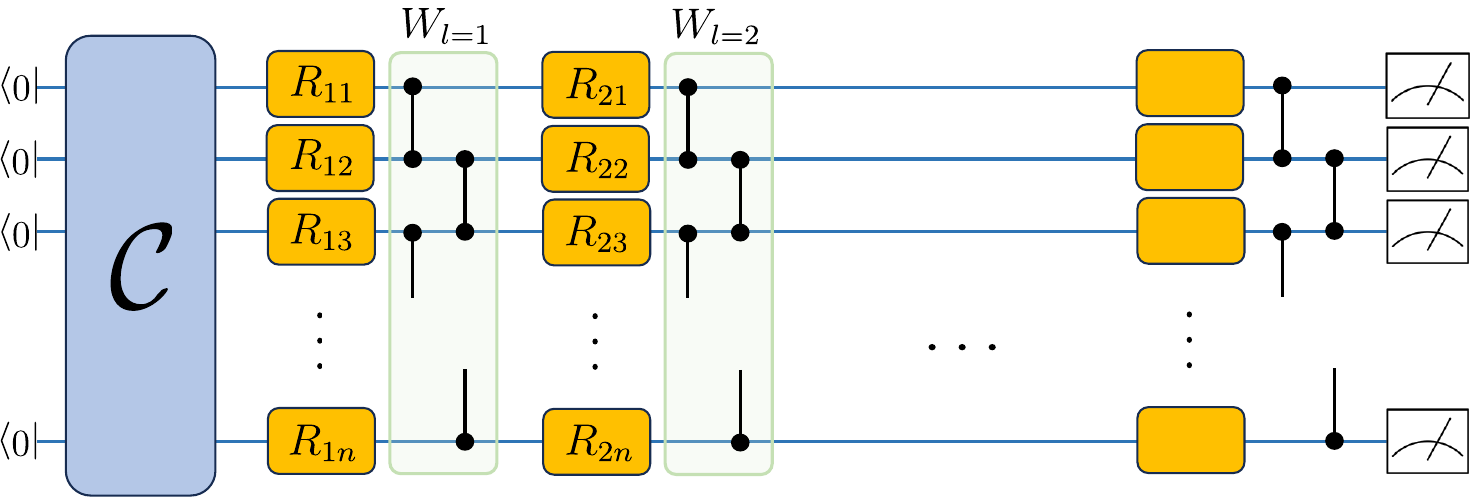}
    \caption{The detailed circuit design of the variational circuit presented in this paper. The data is encoded into the circuit through amplitude encoding to minimize the requirement on the number of qubits and conserve memory for simulation. The parameterized unitary consists of multiple layers of unitaries, with each layer containing a sequence of single-qubit rotations with parameterized rotation axes and angles, denoted as $R_{lm}$ as shorthand for $R_{m {n}_{lm}}(\theta_{lm})$, followed by a sequence of entangling Controlled-$Z$ gates without trainable parameters. An error channel $\mathcal{E}$ is added after every ideal gate $R_{lm}$, constitutes $\Tilde {R}_{lm}$. The inferred probabilities for each potential label are evaluated by measuring the Pauli-$Z$ expected values on each of the qubits. This figure is adapted from Ref. \cite{haiyue_partial_QEC}}
    \label{fig: VQC}
\end{figure}

To evaluate this overhead in a quantum neural network, we numerically estimate the scaling of the spacetime complexities of the full variational circuit with 10 qubits, estimated through the Azure quantum resource estimator \cite{Azure_quantum_resource_estimator, Azure_quantum_resource_estimator_paper}, which is based on Ref. \cite{surface_codes}, as shown in \Cref{tab: full VQC costs}.
\begin{backgroundinformation}{Paramatrised variational circuit}
    A parametrized variational circuit is one critical component of the workflow of quantum neural networks or variational algorithms. The circuit structure used for the spacetime overhead benchmark is described as follows:

    First, the feature of the input data $\bm{x}_i$ is encoded into some quantum state $\rho_{\bm{x}_i}=\mathcal{C}(\bm{x}_i)\ket{0}^{\otimes n}\coloneqq\ket{\psi(\bm{x}_i)}$. For simplicity, here we ignore the cost of the embedding stage by assuming the circuit is insignificant compared to the inference stage. 

    Second, the state is evolved through a non-trivial unitary $U(\bm{\theta})$ with parameters $\bm{\theta}\in \left[-\pi,\pi\right]^D$, which is decomposed into a sequence of $D$ unitaries with parameter $\theta_{lm}$ at layer $l$ on qubit $m$
    \begin{equation}
    U(\bm{\theta})=\prod\limits_{l}W_{l}\prod\limits_{m}e^{-i\theta_{lm}H_{lm}},
    \end{equation}
    where $\theta_{lm}$ and $H_{lm}=\hat{\bm{n}}_{lm}\cdot\bm{\sigma}$, $\bm{\sigma}=(X,Y,Z)^{T}$ are the training parameters for the single-qubit rotation gates $R_{lm}\coloneqq e^{-i\theta_{lm}H_{lm}}$, and $W_l$ is the entangling layer without parameters. For every $R_{lm}$, there are three degrees of freedom, which can be encoded into the Euler angles ($\alpha, \beta, \gamma$), 
    \begin{equation}\label{eq: euler angle decomposition}
    R_{lm}=R_{Z}(\alpha_{lm})R_{Y}(\beta_{lm})R_{Z}(\gamma_{lm}).
    \end{equation}
    The circuit design is illustrated in \Cref{fig: VQC}.
    
    At the end of the circuit, to reduce the impact of ordinary barren plateaus \cite{local_cost_function_1}, the system is measured with respect to the single-qubit Pauli-$Z$ observables.
\end{backgroundinformation}
With the assistance of Microsoft Azure quantum resource estimator \cite{Azure_quantum_resource_estimator, Azure_quantum_resource_estimator_paper} based on defect-based logical qubits introduced in Ref. \cite{surface_codes}, we simulate the total number of physical qubits and the runtime needed to implement a full variational circuit with or without distillations. Although this only demonstrates the behavior for defect-type surface codes, the logic still applies to other surface codes, such as the patch-type ones.
Specifically, given a physical gate error rate of $10^{-3}$, which includes both the physical Clifford gates and the raw magic state, the costs to achieve a full-circuit error budget of $10^{-3}$ and $10^{-4}$ are summarized as in \Cref{tab: full VQC costs}. We note that the full-circuit error budget $\epsilon$ is equal to
\begin{equation}
    \epsilon=\epsilon_{\text{log}}+\epsilon_{\text{dis}}+\epsilon_{\text{syn}},
\end{equation}
for contributions from logical, distillation, and synthesizing errors (the error of synthesizing rotation gates with arbitrary angles), which are equally distributed in our simulation. As a result, the required average logical error rate per qubit per stabilizer cycle $\epsilon_L$ or distilled $T$ gate error rate $\epsilon_T$ must be such that it does not exceed its contribution to the error budget,
\begin{equation}\label{eq: error budget to logical error rate}
    1-(1-\epsilon_{L})^{N_L}\le\epsilon_{\text{log}}\text{ or } 1-(1-\epsilon_{T})^{N_T}\le\epsilon_{\text{dis}},
\end{equation}
where $N_T$ is the corresponding number of logical $T$ gates, $N_L$ total number of patch-cycles. In the limit where $\epsilon_L\ll1$, \Cref{eq: error budget to logical error rate} approximates to 
\begin{equation}
    N_L\epsilon_L\le \epsilon_{\text{log}}\text{ or }  N_T\epsilon_T\le \epsilon_{\text{dis}}.
\end{equation}
As a result, the maximum allowed logical qubit error rate per logical qubit per cycle is
\begin{equation}
    \max_d\left(\epsilon_L=0.03\left(\frac{p}{p^{*}}\right)^{\frac{d+1}{2}}\right)
\end{equation}
that also satisfies \Cref{eq: error budget to logical error rate}.  We omit the next lower integer on $\lfloor\frac{d+1}{2}\rfloor$ as we always assume $d$ is odd. $p$ is the maximum physical error rate per syndrome round, $p^{*}=0.01$ is the physical error rate threshold below which the error rate of the logical qubit is less than the error rate of the physical qubit, and 0.03 is the coefficient extracted numerically from simulations when fitting an exponential curve to model the relationship between the logical and physical error rate. We remark that, due to the discretized nature of the allowed logical error rate per qubit per cycle, the actual full-circuit failure probability is always less than the benchmarked error budget. 

\begin{table}[t!]
    \centering
\begin{tabular}{|c|c|c|c|}
  \hline
  \textbf{Error budget} 
    & \textbf{QVC layers} 
    & \textbf{logical qubit error rate (per cycle)}
    & \textbf{distilled $T$ error rate} \\ 
  \hline\hline
  $1\times10^{-3}$ & 50  & $3.00\times10^{-10}$ & $2.47\times10^{-9}$ \\ 
  \hline
  $1\times10^{-3}$ & 100 & $3.00\times10^{-10}$ & $2.47\times10^{-9}$ \\ 
  \hline
  $1\times10^{-4}$ & 50  & $3.00\times10^{-11}$ & $5.51\times10^{-10}$ \\ 
  \hline
  $1\times10^{-4}$ & 100 & $3.00\times10^{-11}$ & $5.51\times10^{-10}$ \\ 
  \hline
\end{tabular}
\begin{tabular}{|c|c|c|c|c|c|c|}
  \hline
  \multicolumn{4}{|c|}{\textbf{Spatial Cost}} 
    & \multicolumn{3}{c|}{\textbf{Temporal Cost}} \\ 
  \hline
  \shortstack{Code\\distance}
    & \shortstack{Data\\qubits}
    & \shortstack{$T$ factory\\qubits}
    & \shortstack{$T$ factory qubits\\(no distillation)}
    & \shortstack{One logical\\cycle ($\mu$s)}
    & \shortstack{No.\ logical\\cycles}
    & \shortstack{Total\\runtime (ms)} \\ 
  \hline\hline
  15 & 13\,500 & $1.746\times10^6$ & $\sim1.35\times10^4$ & 6.0 & 4\,060 & 24 \\ 
  \hline
  15 & 13\,500 & $1.782\times10^6$ & $\sim1.35\times10^4$ & 6.0 & 8\,410 & 50 \\ 
  \hline
  17 & 17\,340 & $1.746\times10^6$ & $\sim1.73\times10^4$ & 6.8 & 4\,360 & 30 \\ 
  \hline
  17 & 17\,340 & $1.764\times10^6$ & $\sim1.73\times10^4$ & 6.8 & 8\,710 & 59 \\ 
  \hline
\end{tabular}

    \caption{Summary on the cost of the full variational circuit with 10 algebraic qubits ($Q_{\text{alg}}=10$) with different numbers of layers and logical qubits with or without distillation, evaluated by the Azure quantum resource estimator \cite{Azure_quantum_resource_estimator, Azure_quantum_resource_estimator_paper}. For the situation that involves distillations, the figures are calculated based on a physical gate error rate of $10^{-3}$ for both the Clifford gates and $T$ gates before distillation, while achieving a specific logical error rate for the whole variational circuit. For the situation without distillation, the cost of the $T$ factory qubits is simply a number at the same level to the number of data qubits, since they share the same code distance, and there are a maximum of $Q_{\text{alg}}$ logical $\ket{T}$ states to be injected in parallel. It is easy to see that the total overhead differs by almost 2 orders of magnitude between with and without distillation. To a large extent, the time cost is linear with respect to the number of QVC layers. This should still be the case without distillation, yet with a significantly smaller time cost per logical patch-cycle. This table is reproduced from Ref. \cite{haiyue_partial_QEC} with permission.}
    \label{tab: full VQC costs}
\end{table}

Through a careful analysis, we find that to simply execute a 10-logical qubit QVC circuit, millions of physical qubits are required in the $T$-state factory with code distance greater than 15 to achieve a reasonable full-circuit logical error rate, which is two orders of magnitude larger than without distillation. Specifically, for a 100-layer QVC, the cost skyrockets to $1.76\times10^{6}$ physical qubits ($6090d^2$ qubits with code distance 17) in order to achieve a reasonable whole-circuit error budget of $10^{-4}$ with a physical gate error rate of $10^{-3}$. However, without distillation, it is expected that only $\sim60d^2$ qubits are needed, which is comparable to the size of data qubits, for the $T$ factory; therefore, maximally $\sim120d^2$ qubits in total are necessary. However, the time cost should still scale linearly with respect to the number of layers in the circuit, but with a significantly smaller time cost per patch-cycle, which is the number of logical patches times the logical cycles. On the other hand, the time cost is largely a linear relationship with respect to the number of QVC layers. This should still be the case without distillation, yet with a significantly smaller time cost per logical patch-cycle. Overall, it is easy to see that the total overhead differs by almost two orders of magnitude with and without distillation.


\section{Partial Quantum Error Correction}
\begin{figure}[b]
    \centering
    \includegraphics[width=1\linewidth]{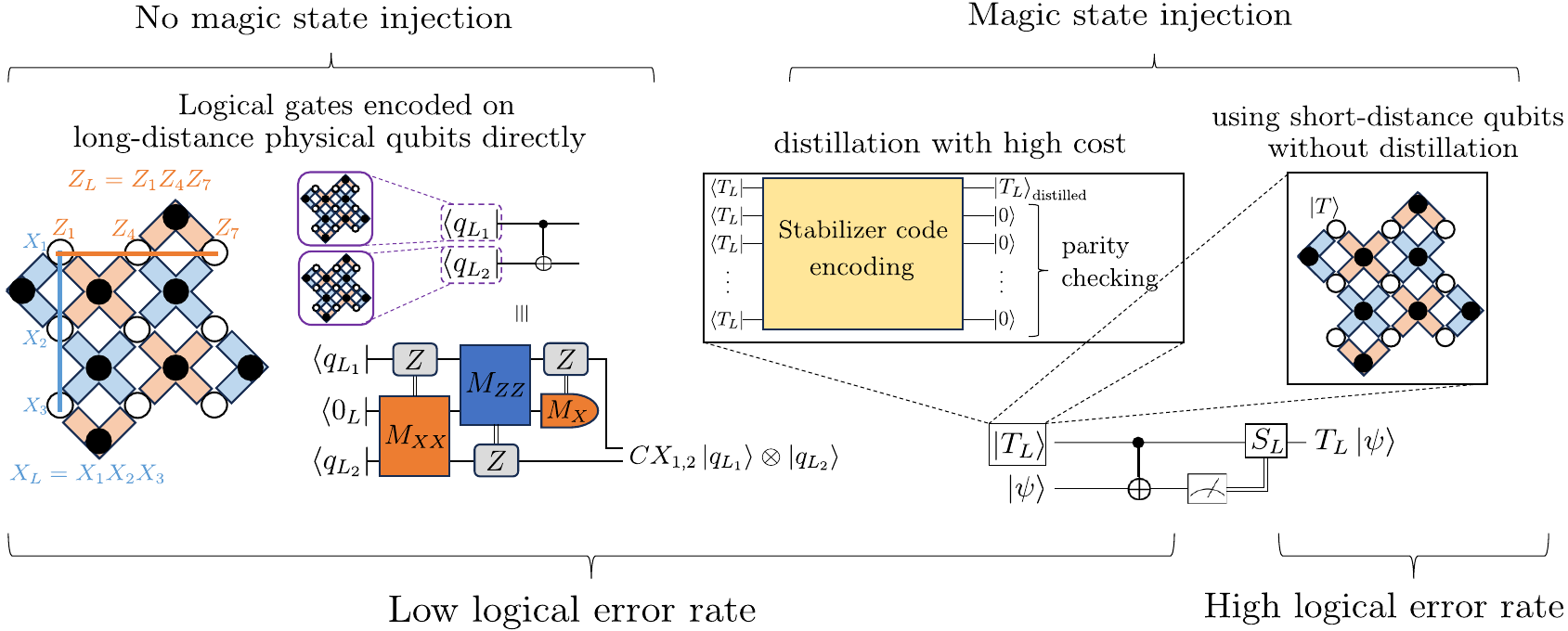}
    \caption{Demonstration of how logical operators are implemented on patch-based surface codes introduced in Ref. \cite{surface_codes}. White dots indicate data qubits, black dots indicate syndrome extraction qubits with orange and blue strips representing Pauli-$X$ and $Z$ stabilizers, respectively. For operators induced from the Clifford group, including $X$, $Z$ and $CX$ gates, their logical operators can be encoded from many physical operators directly, without the need for ancilla qubits \cite{surface_codes, qec_lattice_surgery}. For operators outside the Clifford group, such as the $T$ gate, its logical operator $T_L$ cannot be implemented directly, but must be teleported from an ancilla logical qubit in the state $\ket{T_L}=\frac{1}{\sqrt{2}}(\ket{0_L}+e^{i\pi/4}\ket{1_L})$ via magic state injection. It turns out that $\ket{T_L}$ must be prepared from a single, physical qubit $\ket{T}$ state first, and then perform stabilizer measurements \cite{qec_lattice_surgery}. 
    Or, one could choose to carry out magic state distillation with very high spacetime cost. (b) Without the redundancy of encoding one logical operator from multiple physical operators, the logical error rate for $T$ gates is comparable to the physical $T$ error rate. However, if the state is distilled properly, the logical gate error rate can be suppressed to the same level as other Clifford gates. This figure is adapted from Ref. \cite{haiyue_partial_QEC}.}
    \label{fig: QEC_cost}
\end{figure}
Since depolarizing noise maps to the image of the maximally mixed state, which loses all the quantum information in the long term, its mitigation cannot be accomplished through any self-adaptive mid-circuit manipulations such as dynamical decoupling \cite{dynamical_decoupling}, Pauli twirling \cite{pauli_twirling_1}, whereas other strategies, including zero-noise extrapolation \cite{zero_noise_extrapolation}, probabilistic error cancellation \cite{pauli_twirling_2}, usually require noise parameter characterizations of substantial overhead. Therefore, the necessity of quantum error correction for the quantum variational algorithm becomes apparent when dealing with the noise-induced barren plateaus of quantum neural network training. 

However, we also want to avoid such an overhead of a full-scale QEC that restricts its implementation on near-term quantum processors as discussed in the previous subsection, but still preserving most of its robustness against the hardware noise and its induced barren plateaus. By taking a step back from a fully fault-tolerant surface code-based QEC, we propose a novel approach that relies on partial QEC. In our scheme, Clifford gates (CNOTs) are still kept to be error-corrected, but the $T$ gates are left without any magic state distillation before their injection. By doing this, we can treat the logical variational circuit having its two-qubit gates, as long as they were induced from the Clifford group, with a negligible error rate. However, significant error rates remain for a general single-qubit unitary $U_{\hat{\bm{n}}}(\theta)$ after decomposing it into Clifford plus $T$ gates with error rate $\sim1-(1-\epsilon_T)^{\log_2(1/\epsilon)}$, where $\epsilon_T$ is the raw $T$ gate error rate and the power $\log_2(1/\epsilon)$ is the number of $T$ gates involved in achieving precision $\epsilon$ \cite{T_consumption_for_unitary_improved} of the rotation angle $\theta$. An illustration of the physical implementation of partial-QEC is shown in \Cref{fig: QEC_cost}.

Partial-QEC eliminates the major cost of QEC associated with non-Clifford operations, thereby reducing resource requirements from millions of qubits to merely a few thousand qubits. Our scheme is motivated by the fact that single-qubit parametrized gates are being trained, and therefore, any noise-induced disturbances will be overcome by the training process itself. This is partially inspired by classical machine learning studies, where it has recently been shown that neural networks can be trained regardless of noisy neurons \cite{noisy_classical_CNN,stochastic_classical_NN}. In later paragraphs, it demonstrates that QVC models can be trained with error-corrected CNOTs and noisy single-qubit gates without encountering any trainability issues and without compromising final classification accuracy.  

\subsection{Trainability of QVC with partial-QEC}
To demonstrate the working of our scheme, we selected the 10-class MNIST dataset \cite{MNIST}, which is a standard benchmark in the literature for classical and quantum machine learning models. We experimentally demonstrate the trainability of the QVC75 model (a model with 75 variational layers) for a single-qubit gate error rate comparable to that of state-of-the-art quantum computers, which is reinforced by gradient values above the shot noise. Our work can be generalized to deeper QVC models to further improve classification accuracies, as well as to novel architectures with provable trainability in the presence of hardware noise.

Assuming errors in a general parameterized single-qubit gate persist, and that two-qubit gates are error corrected with a negligible logical error rate, we introduce a depolarizing noise channel after every single-qubit unitary channel $\Lambda_i=R_{\hat{\bm{n}}_{lm}}(\theta_{lm})\bigcdot R^{\dagger}_{\hat{\bm{n}}_{lm}}(\theta_{lm})$ on qubit $i$ with strength $p$. The combined channel is expressed as $\Tilde{\Lambda_i}\coloneqq \mathcal{E}_i\circ \Lambda_i$,
where
\begin{equation}\label{eq: depol channel kraus rep}
\begin{aligned}
    \mathcal{E}_i(\rho)&=(1-p)\rho+\frac{p}{3}\left(X_i\rho X_i+Y_i\rho Y_i+Z_i\rho Z_i\right),
    \\
    \Lambda_i(\rho)&=U_i\rho U_i^{\dagger}
\end{aligned}
\end{equation}
are the noise channel and unitary channel in Kraus representation, respectively. Since the noise channel acts on the logical space, $p$ is interpreted as the combined logical $X$, $Y$, and $Z$-flip error rate. In practice, the channel would map the density operator after every decomposed gate of $R_{\hat{\bm{n}}_{lm}}$. However, since any unitary commutes with the depolarizing channel, it is convenient to treat it as in \Cref{eq: depol channel kraus rep}.

To evaluate the trainability of the variational circuit with only single-qubit noise, we used a learning rate of 0.005 and a total of 15000 images split into 50 images per batch in our training set, and 250 images in our test set, which are drawn and ordered randomly. The expectation values are evaluated on the Pauli-$Z$ operators, and every circuit is executed with 10000 shots to simulate real hardware with non-zero shot noise and avoid potential saddle points \cite{shot_noise_avoids_saddlepoints}. 

\begin{figure}[t!]
    \centering
    \subfloat{
    \includegraphics[width=.484\linewidth]{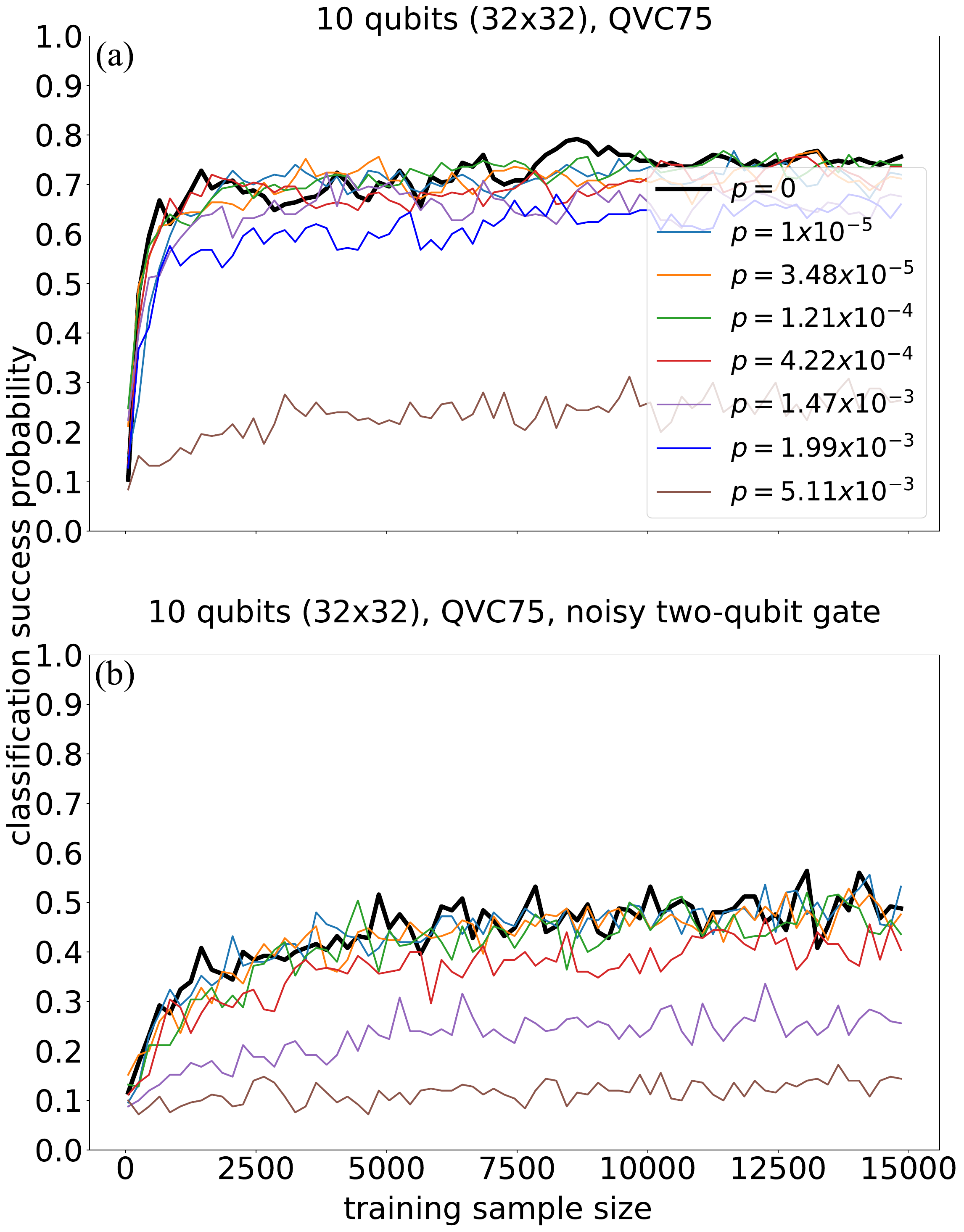}}
    \subfloat{\includegraphics[width=.516\linewidth]{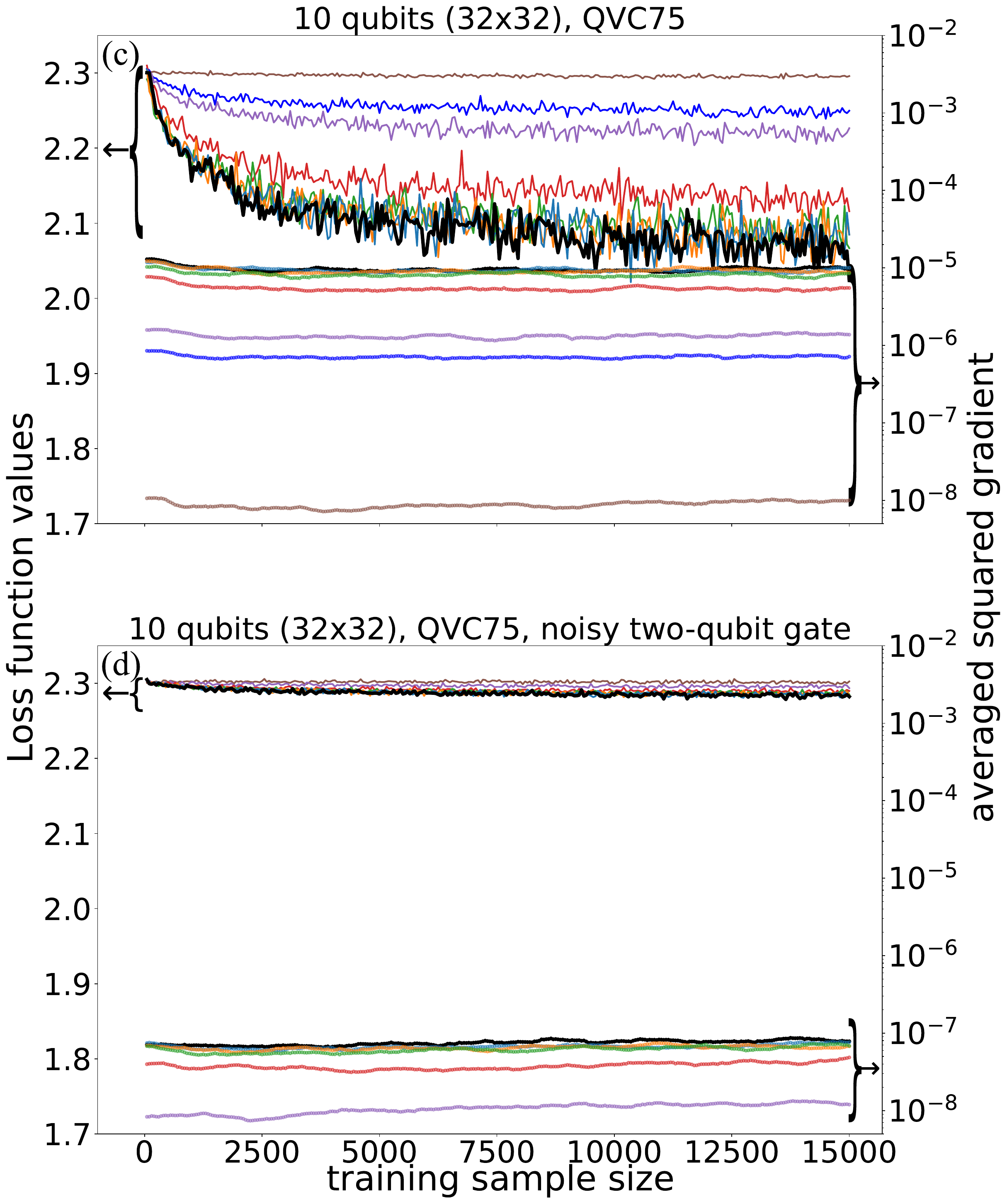}}
    \caption{The performance of the QVC in MNIST number classification problems with (a) 75 quantum layers but without a classical layer, and (b) 75 layers with a fully-connected classical layer and noisy two-qubit gate. The classification success rates inferred from the test datasets are plotted against the total number of images trained. The noise-free simulation is highlighted with black, thickened lines to contrast with noisy ones. Since we employ amplitude encoding, the number of pixels per image is exactly $2^n$, which corresponds to the dimension of $\sqrt{2^{n}}\cross \sqrt{2^{n}}$, where $n$ is the number of qubits for the variational circuit. The legend denotes the strengths of the depolarizing channel from 0 to $5.11\cross 10^{-3}$. (c) The corresponding cost function values (left axis) and their average gradients squared (right axis), $\mathbb{E}_k$$\left((\nabla_{\theta_k}\mathcal{L})^2\right)$, of the trainable parameters (classical and quantum), where the arrows indicate the axis correspondence of the plots. A clear trend of flattening loss landscapes and vanishing gradients can be observed as the depolarizing strength increases. When the gradients drop to a scale around $10^{-8}$, i.e. $p_{\text{depol}}=5.11\times10^{-3}$, due to shot noise, the model becomes distinctively not trainable. (d) For the noisy two-qubit gate plot where the classical layer is included, the averaged cost function gradients with respect to all parameters, including both the quantum and classical layers, actually increase after being trained with more images. In contrast, the classical layer gradients by themselves are behaving as expected, as shown in the zoomed figure. This figure is reproduced from Ref. \cite{haiyue_crosstalk} with permission.}
    \label{fig: depol plots long layers}
\end{figure}

\begin{figure}[t]
    \centering
    \includegraphics[width=.6\linewidth]{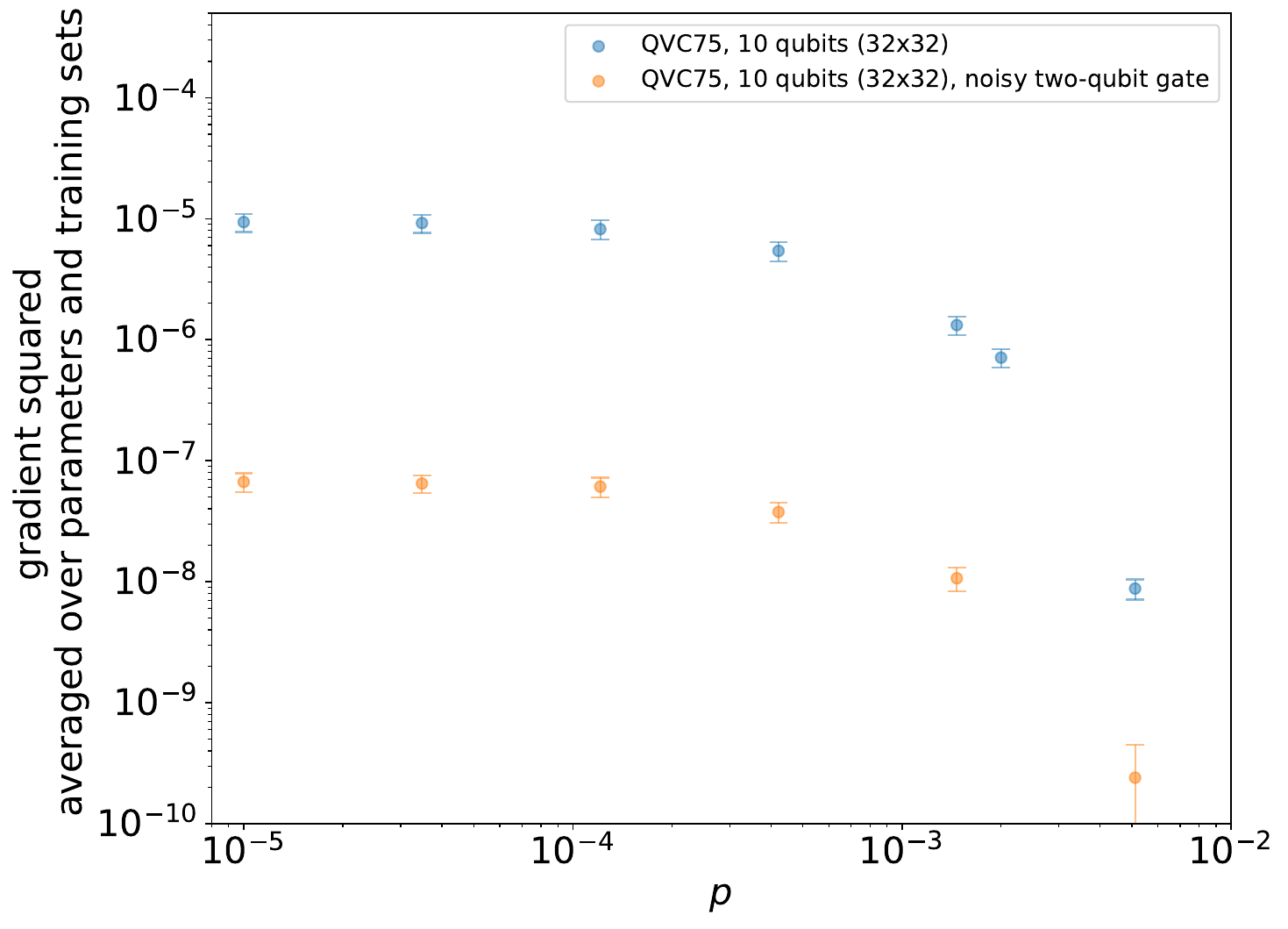}
    \caption{The gradients squared averaged over all trainable parameters and all iterations in the training versus the depolarizing strength $p$ in logarithmic scale. Error bar takes the standard deviation of the mean average among all iterations. This figure is reproduced from Ref. \cite{haiyue_crosstalk} with permission.}
    \label{fig: gradients_vs_p_long_layers}
\end{figure}

As shown in \Cref{fig: depol plots long layers}a, the classification success rate attains a lower saturation more quickly when the depolarization strength increases. There is also a clear trend that both the gradients and the cost function landscape quickly flatten as the depolarizing noise becomes stronger, as is evident in \Cref{fig: depol plots long layers}c and \Cref{fig: gradients_vs_p_long_layers}. These results agree with the prediction of \cite{noise_induced_barren_plateaus}, reinforcing the argument that depolarizing noise induces barren plateaus, which have vanishing gradients that scale exponentially with their strength. 

Most importantly, as shown by the blue curves in \Cref{fig: depol plots long layers}a and \Cref{fig: depol plots long layers}c, despite shallower gradients and a smaller asymptotic success rate, the model is still trainable and above the shot noise at least for depolarizing strength of $p=1.99\cross 10^{-3}$ with only a slight reduction in the classification success rate. This corresponds to a logical parametrized single-qubit gate $U_{\hat{\bm{n}}_{lm}}(\theta_{lm})$ with net error rate of $1.33\times 10^{-3}$, a value can be achieved with underlying raw logical $T$ gate error rate no more than $\epsilon_T=10^{-4}$ by decomposing $U_{\hat{\bm{n}}_{lm}}(\theta_{lm})$ into Clifford + $T$.
\begin{backgroundinformation}{Depolarizing noise strength and logical $T$ gate error rate analysis}\label{appendix: Depolarising noise strength and actual gate error analysis}

Given a depolarizing channel with a depolarizing noise strength $p_{\text{depol}}$ added after an arbitrary single-qubit unitary $U$, the actual decay rate $p$ is determined from the standard randomized benchmarking \cite{randomised_benchmarking}. Specifically,
\begin{equation}
    p\coloneqq \int_{}d\psi\text{Tr}(\ket{\psi}\bra{\psi} \Tilde{U}^{\dagger}\circ\Tilde{U}(\ket{\psi}\bra{\psi}))
\end{equation}
where $\Tilde{U}$ and $\Tilde{U}^{\dagger}$ are the channels of the noisy $U$ and $U^{\dagger}$ gates. Since the depolarizing channel commutes with any single-qubit unitaries, let $\ket{\psi}=V_{\psi}\ket{0}$, and absorb $V_{\psi}$ into $U$ as $W_{\psi}=UV_{\psi}$, we have
\begin{equation}\label{eq:randomised_benchmarking_error_rate_for_depolarising_channel}
\begin{aligned}
    p
    &=\int d\psi\text{Tr}\{V_{\psi}\ket{0}\bra{0}V_{\psi}^{\dagger} U^{\dagger}\mathcal{E}^2_{\text{depol}}(UV_{\psi}\ket{0}\bra{0}V_{\psi}^{\dagger}U^{\dagger})U\}\\
    &=\int d\psi\text{Tr}\{W_{\psi}\ket{0}\bra{0}W_{\psi}^{\dagger}\mathcal{E}^2_{\text{depol}}(W_{\psi}\ket{0}\bra{0}W_{\psi}^{\dagger})\}\\
    &=\text{Tr}\{\ket{0}\bra{0}\mathcal{E}^2_{\text{depol}}(\ket{0}\bra{0})\}.
\end{aligned}
\end{equation}
Substituting $\mathcal{E}_{\text{depol}}\left(\frac{I+Z}{2}\right)=\frac{I+\left(1-\frac{4p_{\text{depol}}}{3}\right)Z}{2}$ into \Cref{eq:randomised_benchmarking_error_rate_for_depolarising_channel}, it is apparent that
\begin{equation}
    p=\frac{1+\left(1-\frac{4p_{\text{depol}}}{3}\right)^2}{2}.
\end{equation}
Therefore, the actual gate error rate is given by
\begin{equation}
    r\coloneqq 1-p - (1-p)/d,
\end{equation}
where $d$ is the dimension of the Hilbert space, for a single-qubit state $d=2$. However, such an error rate for an arbitrary single-qubit unitary must be shared by a sequence of Clifford + $T$ gates in the practice of error correction code, and the number of $T$ gates required scales as $\sim1.5\log_2(1/\epsilon)$ for a precision of $\epsilon$ \cite{T_consumption_for_unitary}. Therefore, based on the assumption of partial QEC that only considers the significant error rate for $T$ gates, the relationship between the logical $T$ gate error rate $\epsilon_T$ to the arbitrary rotation is given as
\begin{equation}
    r\sim1-(1-\epsilon_T)^{1.5\log_2(1/\epsilon)}.
\end{equation}
In literature \cite{chamberland2020very,magic_state_distillation_cost1,magic_state_distillation_cost2}, a physical error rate of $10^{-4}$ was often used, which arises from the expectation that it has a reasonably good error suppression factor, which scales as $ O\left(\frac{p_{\text{threshold}}}{p}\right)$, and practical reachability. Since the partial QEC protocol would still require the physical gate error rate to be sufficiently below the threshold around $10^{-2}$ \cite{surface_codes}, at the time when full-scale QEC becomes feasible, a physical two-qubit error rate of $10^{-4}$ is fundamental and achievable, which is an order of magnitude lower than what has been achieved nowadays. According to the estimation in \cite{physical_T_and_logical_T_error_rate}, which uses the SI1000 error model \cite{SI1000} by assuming that $p_2=10p_1$, we have $\frac{2}{3}p_2\le\epsilon_T\le p_2$ for $p_2$ sufficiently small, where $p_1$, $p_2$ are the physical single and two-qubit gate error rates, respectively. Therefore, we use $\epsilon_T=10^{-4}$ as the upper bound for $p_2=10^{-4}$. For a logical $T$ gate with an raw error rate of $\epsilon_T=10^{-4}$, the final $U$ gate error rate can be derived as $r=1.99\cross 10^{-3}$ 
with a small tolerance of $\epsilon=10^{-4}$
in its rotation angle. This implies that the corresponding depolarizing channel strength is $p_{\text{depol}}=2.99\cross 10^{-3}$.

In one of the latest literature \cite{T_consumption_for_unitary_improved}, the authors improve the scaling of the number of $T$ gates required to be $\sim \log_2(1/\epsilon)$ by introducing $T_X$ gates that are $\pi/4$ rotations in $X$ direction, but have the same cost as normal $T$ gates. In this case, the actual gate error can be further relaxed to $1.33\cross 10^{-3}$,
which corresponds to $p_{\text{depol}}=1.99\times10^{-3}$.
At this rate, we show that the model is still trainable in \Cref{fig: depol plots long layers}, where the classification rate is only slightly decreased.
\end{backgroundinformation}

Furthermore, as shown in \Cref{fig: depol plots long layers}b and \Cref{fig: depol plots long layers}d, we test the QVC model performance when the two-qubit gate noise is also turned on, which is aimed at simulating when the circuit is purely executed on physical qubits without error-correction, thereby demonstrating the necessity of partial QEC in balancing the model's trainability and overhead compared to no QEC. Similarly to the single-qubit gate noise channel, the two-qubit gate noise is added after the noise-free unitary: $\Tilde{\Lambda_i}\coloneqq \mathcal{E}_{(q_1,q_2)}\circ \Lambda_{(q_a,q_b)}$, where $q_1$, $q_2$ are the two qubits the gate is acting on. The overall noise channel $\mathcal{E}_{(q_1,q_2)}$ contains two subchannels: depolarizing channel $\mathcal{E}_{\text{depol}_i}$ on qubit $i$ with the same structure as in \Cref{eq: depol channel kraus rep}, and crosstalk nearest-neighbour $ZZ$ couplings $\mathcal{E}_{\text{crosstalk}_{(q_1,q_2)}}$ on qubits $q_1$ and $q_2$, where
\begin{equation}
    \mathcal{E}_{\text{crosstalk}_{(q_1,q_2)}}[\bigcdot] = U_{Z_{q_1}Z_{q_2}}\bigcdot U^{\dagger}_{Z_{q_1}Z_{q_2}}.
\end{equation}
The unitary $U_{Z_{q_1}Z_{q_2}} = \exp\left(-i\alpha Z_{q_1} Z_{q_2}\right)$ is simply a two-qubit unitary with generating Hamiltonian $ZZ$ and angle of rotation $2\alpha$. Since multiple channels are involved, we describe each channel $\mathcal{E}_i \coloneqq e^{\mathcal{L}_i}$ with Lindbladian generator $\mathcal{L}_i$ in the superoperator representation, where
\begin{equation}
    \mathcal{E}_i=\sum\limits_{j}K_{ij}^{T}\otimes K_{ij},
\end{equation}
$K_{ij}$ are the Kraus operators of channel $i$. Therefore, the net Lindbladian
\begin{equation}
\begin{aligned}
\mathcal{L}&=\mathcal{L}_{\text{depol}_{q_1}} + \mathcal{L}_{\text{depol}_{q_2}} \\
    &+ \mathcal{L}_{\text{crosstalk}_{(q_1-1,q_1)}} + \mathcal{L}_{\text{crosstalk}_{(q_2,q_2+1)}}
\end{aligned}
\end{equation}
is the sum of the Lindbladian generator of all noise channels for the master equation,
\begin{equation}
    \frac{d\rho}{dt}=\mathcal{L}\left[\rho\right],
\end{equation}
and have the solution $\vert\rho(t)\rrangle= e^{\mathcal{L}t}\vert\rho(0)\rrangle$, where $\vert\rho\rrangle$ denotes the density operator in vectorized form.
For a more realistic simulation, the depolarizing and crosstalk noise strengths are characterized by the IBM superconducting quantum computer $ibm\_torino$.

When two-qubit gate noises are added back, the classification success rate is much worse. In all situations, the classification success rate never cross 0.5 and the average squared gradient  $\mathbb{E}_k$$\left((\nabla_{\theta_k}\mathcal{L})^2\right)$ is below $10^{-7}$. At the single-qubit depolarizing channel strength of $p=1.47\cross 10^{-3}$, classification success rate barely increases and the model is almost non-trainable, whereas the model behaves properly for the case without the two-qubit gate noises. The comparison between noisy and noise-free two-qubit gates demonstrates the utility of partial QEC, which significantly improves the trainability of the machine learning model while still achieving a drastic overhead reduction by forgoing distillations.

\section{Implementation of Quantum Error Detection with Quantum Machine Learning}
In addition to the challenges associated with implementing QEC with QML on physical hardware, there are challenges associated with implementing QEC with QML in classical simulations due to the high computational overhead. For this reason, QML research has largely ignored the impact of noise, or focused on error mitigation rather than full quantum error correction. To address this research gap, we investigated the performance of a simple, parity-classifying QVC implemented with the [[4,2,2]] error-detecting stabiliser code within simulated noisy environments. To our knowledge, this work constitutes the first study on the implementation of a QML algorithm with a QEC code. Although our study was limited to error detection only, its results are generalisable to fully error-correcting codes and more complex QML algorithms. 

\subsection{Logical Encoding}\label{sec:logical_encoding}
\begin{figure*}
\centering
\begin{quantikz}
\ket{0} & \phantomgate{0} & \phantomgate{0}\gategroup[2,steps=1,style={dashed, rounded corners, inner sep=6pt}]{Basis Encoding} & \phantomgate{0} & \gate{R_{X}(\theta)}\gategroup[2,steps=4,style={dashed,rounded corners,inner sep=6pt}]{Variational Component}  & \gate{R_{Z}(\theta)} & \ctrl{1} & \gate{R_{Y}(\theta)} & \gate{\sigma_z} & \meter{} \\
\ket{0} & \phantomgate{0} & \gate{X} & \phantomgate{0} & \gate{R_{X}(\theta)}& \gate{R_{Z}(\theta)} & \targ{} & \gate{R_{Y}(\theta)} &\phantomgate{0} & \phantomgate{0}
\end{quantikz}
\caption{The Variational Quantum Classifier (VQC) with an example input state of $\ket{01}$ and rotational parameter $\theta$. This figure is adapted from Ref.~\cite{adermann_QEC_VQC}.}
\label{VQC}
\end{figure*}
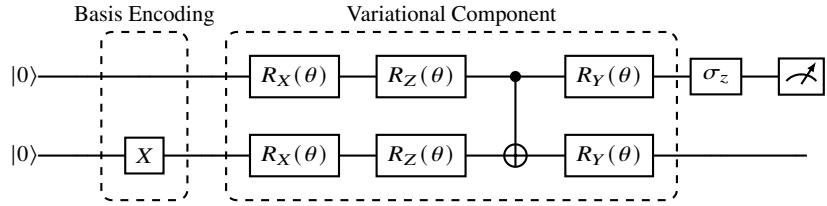

We chose a very simple 2-qubit QVC for our experiments (displayed in Figure~\ref{VQC}), to minimise computational overhead. The QVC takes as input two qubits encoded in the basis encoding, and classifies their parity through measurement of the first qubit in the $Z$ basis. The quantity measured is the expectation value over $1000$ shots. We use only one rotational parameter, $\theta$, to train the classifier, as any more than one leads to overfitting. The classifier is able to reach an accuracy of $1.0$ within 100 training iterations. 

We used the [[4,2,2]] 4-qubit Calderbank-Shor-Steane (CSS) stabiliser code for our error detection, as it is the simplest stabiliser code that protects against $X$ and $Z$ single-qubit errors \cite{Vaidman_1996}. It encodes 2 logical qubits using 4 physical qubits, and can only facilitate detection (not correction) of single-qubit errors. Errors are detected by taking measurements of ancilla qubits after applying stabilisers to the physical qubits, known as syndrome extraction. 

We used the following mapping to encode 2 logical qubits with 4 physical qubits: 
\begin{align}
    \ket{00}_{L} &= \frac{1}{\sqrt{2}}(\ket{0000} + \ket{1111})\\
    \ket{01}_{L} &= \frac{1}{\sqrt{2}}(\ket{0011} + \ket{1100})\\
    \ket{10}_{L} &= \frac{1}{\sqrt{2}}(\ket{0101} + \ket{1010})\\
    \ket{11}_{L} &= \frac{1}{\sqrt{2}}(\ket{0110} + \ket{1001}), 
\end{align}
where the left-hand side represents the four different 2-logical qubit states (indicated by the subscript $L$), and the right-hand side represents the physical qubit states. 

With these definitions, the CNOT gate in the QVC is logically encoded by a SWAP gate between the first two physical qubits. The logical encoding for the rotation gates require additional ancilla qubits (one ancilla per rotation gate to be encoded), where the ancilla qubits undergo the rotation instead of the physical qubits. This process generally requires a series of CNOT gates before and after the application of each rotation gate, which entangle the physical (and logical) qubits with the ancilla qubits. 

In our logical encoding, the quantum state of the full qubit system (including ancillas), is always of the form: 
\begin{equation}
\label{superposition_state}
    \ket{\Psi} = \sum_{i=0}^{3} c_i \ket{\psi_i}_L \bigotimes_{j=0}^{n-1}\,\ket{\phi_{i}}_{a_j},
\end{equation}
where $\ket{\psi_i}_L$ represents each of the four possible logical basis states, $c_i$ is the complex coefficient associated with the $i$-th logical basis state, $\ket{\phi_i}_{a_j}$ represents the state of the $j$-th ancilla qubit, $a_j$, associated with the $i$-th logical basis state, and $n$ is the total number of ancillas that have been introduced into the system. 

To illustrate, the full quantum state after the first two $R_{X}$ rotations is given by: 
\begin{align}
\label{superposition_state}
    \ket{\Psi} &= -i\mathrm{cs}\ket{00}_L \ket{0}_{a_1}\ket{0}_{a_2} + \mathrm{c^{2}}\ket{01}_L\ket{0}_{a_1}\ket{1}_{a_2} \nonumber\\
    &- \mathrm{s^{2}}\ket{10}_L\ket{1}_{a_1}\ket{0}_{a_2} - i\mathrm{sc}\ket{11}_L\ket{1}_{a_1}\ket{1}_{a_2},
\end{align}
where $c$ and $s$ are short-hand notations for cos($\theta$) and sin($\theta$), the logical states are subscripted with $L$ and the ancilla states are subscripted with $a_j$. There are two ancilla states ($a_1$, $a_2$) invoked because two rotations have occurred. By the end of a logical operation, the ancilla qubits in each term reflect the logical state within the same term. 

The steps we used to logically perform the double $R_Y$ and $R_X$ gates are outlined below:
\begin{enumerate}
\item \textbf{New ancilla initiation:} If no rotations have been performed in previous steps, initiate two new ancilla qubits to match the initial input logical state to the circuit. For example, if the initial input logical state is $\ket{01}_L$, then the two new ancillas, $a_j$ and $a_{j+1}$, should respectively be in the states $\ket{0}$ and $\ket{1}$. If double rotations have been performed in previous steps, invoke the new ancilla qubits in the $\ket{0}$ state, then match the new ancilla states to the previous two ancilla states that were invoked to perform the last double rotation, by applying CNOT gates to the new ancilla qubits controlled by the two previous ancilla qubits. 
\item \textbf{Change previous ancilla states:} If double rotations have been performed in previous steps, apply CNOT gates to each of the previously invoked ancilla qubits, controlled by the new ancilla qubits. For example, for newly invoked ancilla qubits $a_j$ and $a_{j+1}$, we apply CNOT gates to any previous ancillas $a_{j-2}$, $a_{j-4}$, ..., $a_0$ and $a_{j-1}$, $a_{j-3}$, ..., $a_1$ controlled by $a_j$ and $a_{j+1}$ respectively. 
\item \textbf{Change logical state:} Apply CNOT gates to physical qubits $q_1$ and $q_3$ controlled by newly initiated ancilla $a_j$, and CNOT gates to $q_2$ and $q_3$ controlled by newly initiated ancilla $a_{j+1}$. 
\item \textbf{Apply rotation gate:} Apply the relevant rotation gates to each of the 2 ancillas. 
\item \textbf{Undo logical state change:} Apply the same set of CNOT operations targeting the physical qubits and controlled by the 2 ancillas as was performed in Step 3. 
\item \textbf{Undo previous ancilla state change:} Apply the same set of CNOT operations as in Step 2, targeting the ancillas invoked for previous rotations, and controlled by the newest ancilla states. 
\end{enumerate}

The steps for implementing the $R_Z$ gates are much simpler and do not require as many CNOT gates between physical and ancilla qubits. We only match the newly introduced ancilla qubits to the previous two ancilla qubits, then apply the $R_Z$ rotation gate to each new ancilla qubit. 
For the logical CNOT gate, we applied additional CNOT gates after the SWAP gate to ensure matching between the ancilla and logical states in each term of the full quantum state of the system. 

The above steps can be used to logically implement single qubit rotations, in which case we only need to introduce one ancilla qubit each time. 
Our approach can also be generalised to circuits with any number of rotation gates. 

The logical circuit does not allow the direct $Z$ basis measurement of the first logical qubit, so we conducted the logical equivalent by measuring the probability distribution across the 16 states spanned by the four physical qubits. Using these probabilities, we calculated the equivalent probability distribution for the four logical 2-qubit states, from which we determined the expectation value of the $Z$ basis measurement of the first logical qubit. 

\subsection{Experimental Simulations}
As stabiliser codes can only detect and correct combinations of $X$ and $Z$ errors, we only consider $X$, $Y$ and $Z$ errors for our noise models, in the form of probabilistic gate noise and depolarising environmental noise. This means that our noise models are inherently unable to capture the full range of noise and errors that might arise in physical NISQ systems. However, since our aim is to evaluate the effectiveness of the [[4,2,2]] stabiliser code in improving training outcomes, we only need to simulate noise that the code is theoretically capable of detecting. 

We implement the gate noise model with single-qubit ``error" gates applied after each single-qubit gate, where there is a probability (or Pauli Error Rate), given by $p$ (with $0 < p < 1$), of either an $X$, $Y$ or $Z$ error occurring. Additionally, we apply single-qubit error gates after each 2-qubit gate, on the same qubits targeted by the 2-qubit gates, doubling the Pauli Error Rate compared to single-qubit gates, to better model the increased error rate for multi-qubit gates compared to single-qubit gates. 

Our environmental noise model is a highly simplified model consisting of $X$,$Y$ and $Z$ errors. We apply Pauli errors at regular intervals to each physical and ancilla qubit at the same time. Applying noise at regular intervals mimics the cumulative build-up of errors in quantum circuits that occur as a result of environmental noise, where the specific regularity of the noise injections reflects the typical relaxation time and dephasing time of the system. Although this model does not include amplitude damping noise, dephasing noise, or the entire span of complex errors that could arise from realistic noise models, it is able to capture a range of alterations that may occur to the qubits as a result of energy loss to the system and decoherence, and is compatible with our choice of QEC code. 

We ran simulations of the logically-encoded circuit under the gate noise and environmental noise models in a classical high-performance computing environment. The application of each stabiliser and syndrome extraction adds one extra qubit to the system, hence the resource overhead for the simulations ranged from 12 qubits to 20 qubits depending on the number of syndrome extraction rounds performed. 

Since there are only four unique data points that can be used to train the VQC (namely, [0,0,0], [0,1,1], [1,0,1] and [1,1,0]), we duplicated the data to produce 24 samples for training the VQC. We used a batch size of 8 and ran the training for 100 iterations each, which was more than sufficient for convergence in a zero noise environment. Since the [[4,2,2]] stabiliser code can only detect errors, we discarded shots where at least one $X$ or $Z$ error was detected. 

We applied the noise models to both ancilla and physical qubits in the system, but kept the syndrome extraction qubits noise-free. We chose a Pauli Error Rate ranging from $0.001$ to $0.01$ for both models, consistent with current NISQ device capabilities~\cite{Arute2019}. For the environmental noise model, we chose a Pauli error injection regularity of once every 4 gates, with the same Pauli Error Rates as used for the gate noise model. 
%

\subsection{Impact of Noise on VQC Training and Performance}
\begin{figure*}
\centering
\includegraphics[trim={0.9cm 0cm 0.9cm 0.9cm}, clip=true, width=\linewidth]{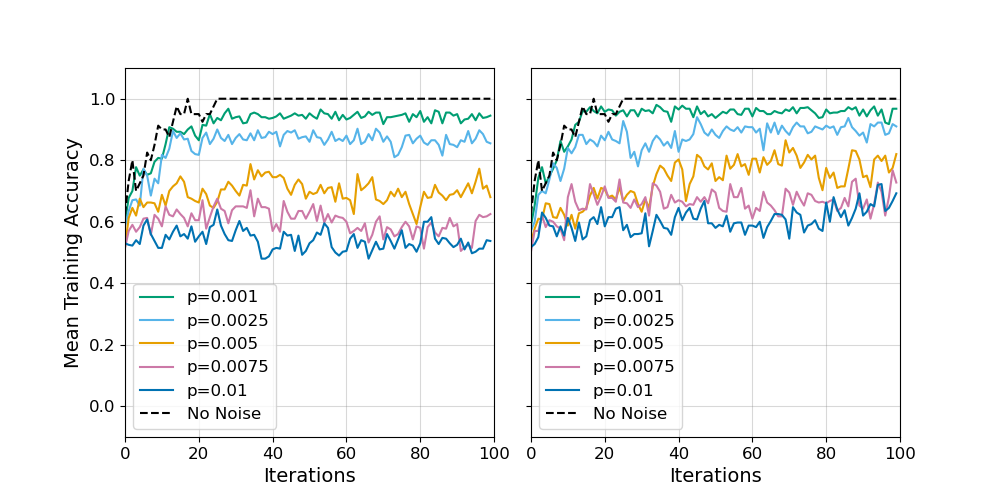}
\caption{\label{noerrordetection} Mean training accuracy of the logically-encoded VQC when training under different levels of gate (left) and environmental (right) noise, ranging from $p=0.001-0.01$. The black dashed line indicates the accuracy obtained from training without noise. This figure is adapted from Ref.~\cite{adermann_QEC_VQC}.}
\end{figure*}

In Figure~\ref{noerrordetection}, we show the evolution of the mean accuracy achieved during training under both noise models, with varying noise levels expressed as Pauli Error Rates and without error detection. The mean was calculated from 10 simulations with different starting seeds. As expected, higher noise levels produce lower mean final training accuracies. The QVC struggles to learn when $p\geq0.005$, which is evident in how the training accuracy stays roughly constant throughout training. When noise levels satisfy $p\leq0.0025$, learning appears hampered but not impossible. 

Our results indicate that the simple QVC model is fairly susceptible to noise, which is likely due to its reliance on only two qubits to learn and make predictions. While more complex QML algorithms generally have greater learning capacities and may be more robust to noise and errors, they also require more qubits and gates, increasing the potential for errors. Consequently, a combination of error correction and mitigation is most likely needed to train and implement both simple and complex QML algorithms on NISQ devices (where noise levels can exceed $p=0.005$). 

\subsection{Effectiveness of Error Detection on QVC Training and Performance}
\begin{figure*}
\includegraphics[trim={0.9cm 0cm 0.9cm 0.9cm}, clip=true, width=\linewidth]{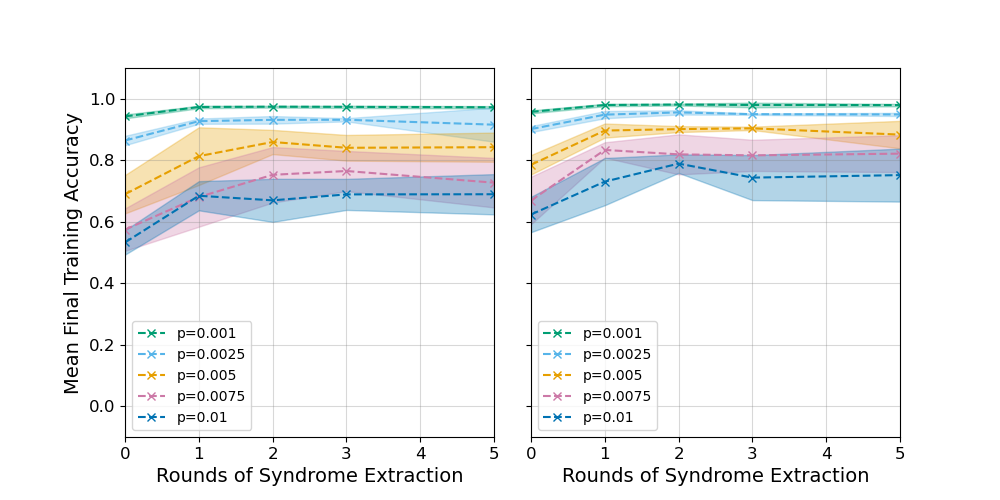}

\caption{\label{witherrordetection} Mean final training accuracy of the logically-encoded VQC under different levels of gate (left) and environmental (right) noise ranging from $p=0.001-0.01$, and with different degrees of error detection implemented from 0 to 5 rounds of syndrome extractions. The mean final training accuracy is calculated from the training accuracies attained by the VQC over the final 40 iterations of training, after convergence. The standard deviation of the mean final training accuracy from 10 training runs is represented by the shaded regions. This figure is adapted from Ref.~\cite{adermann_QEC_VQC}.}
\end{figure*}

In Figure~\ref{witherrordetection}, we display the effect of implementing the [[4,2,2]] stabiliser code with varying numbers of syndrome extraction rounds on the final training accuracy achieved after model convergence. In this and subsequent figures, we are interested in the mean final training accuracy, which we calculated by taking the mean of the accuracies recorded over the last 40 iterations of training (as we can assume the training has stabilised by this point), and averaging this mean over 10 simulations of training. We also report the first standard deviation associated with this mean. 

Under both noise models, we observe that for low noise (which we define as $p \leq 0.0025$), the training accuracy always improves with increasing number of syndrome extraction rounds and there is high consistency in the final training accuracy recorded across the 10 simulations. However, for higher levels of noise (specifically, $0.005 \leq p \leq 0.01$), the final training accuracy is quite variable (large spread in values), and occasionally worsens with more syndrome extractions. 

There is also evidence of limitations in the effectiveness of error detection at low noise levels. We find that even for $p\leq0.0025$, the increase in syndrome extraction rounds produces no significant increase in final training accuracy beyond two extraction rounds. The final accuracy does not reach $1.00$ even at low noise levels and with five rounds of syndrome extractions. 

These results suggest that there is noise in the logically-encoded circuit that syndrome measurements cannot detect. The only possible source of this noise is the ancilla qubits, which we do not apply any syndrome extractions to, but are entangled with the physical qubits via multiple CNOT gates. 

\subsubsection{Ancilla Qubit Noise}
\begin{figure*}
\includegraphics[trim={0.9cm 0cm 0.9cm 0.8cm}, clip=true, width=\linewidth]{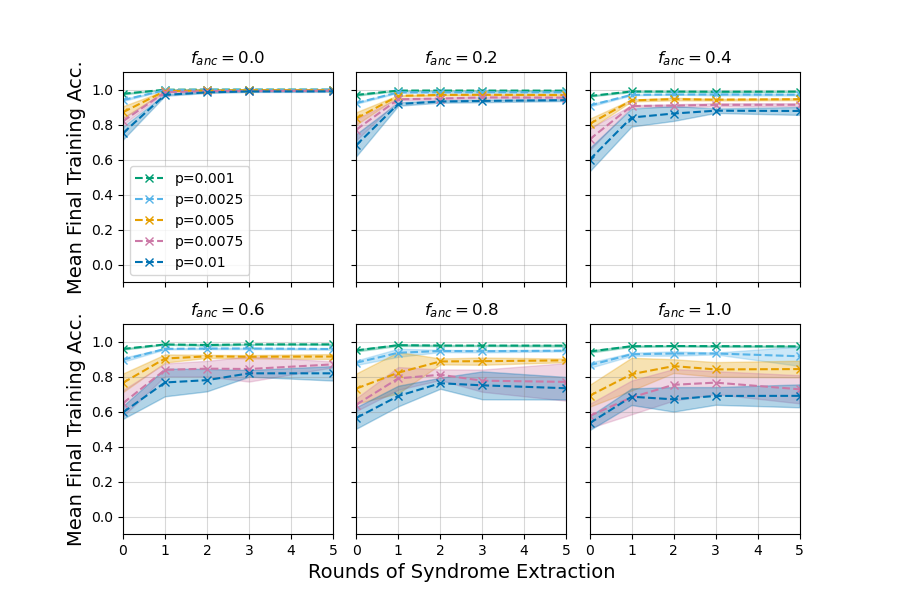}
\caption{\label{GateDP_Plot} Impact of ancilla qubit error rate on the mean final training accuracy, under the gate noise model with physical qubit error probability ranging from $p=0.001-0.01$. Each subplot displays the variation in mean final training accuracy with the number of rounds of syndrome extraction, for a specific combination of ancilla and physical qubit error probabilities. The ancilla error rates are expressed as a fraction of the physical error rates, denoted by the fraction $f_{anc}$. The standard deviation in the meaning final training accuracies, each calculated from 10 training runs, is shown by the shaded regions. This figure is adapted from Ref.~\cite{adermann_QEC_VQC}.}
\end{figure*}

To determine if the ancilla qubit errors are responsible for the limited effectiveness of the [[4,2,2]] stabiliser code in detecting errors, we trained the QVC under different levels of ancilla qubit noise. We show in Figure~\ref{GateDP_Plot} 
the evolution in the mean final training accuracy under the gate noise model, as the number of syndrome extractions increase, and with different levels of ancilla noise. We use $f_{anc}$ to denote the fraction of the physical Pauli Error Rate that we apply to the ancilla qubits. For example, $f_{anc}=0.5$ means that the ancilla Pauli Error Rate is half the physical Pauli Error Rate. As the results under the environmental noise model are very similar, we limit our analysis to the results obtained under the gate noise model only. 

It is clear from the plot that as the ancilla error rate increases, the syndrome extractions become less effective and less reliable, confirming our earlier hypothesis. There is a greater spread in the final training accuracy, as well as an increase in non-monotonicity (which we suggest is a consequence of the greater variance), when the ancilla error rate is higher. Lower noise levels produce better training outcomes than higher noise levels for $f_{anc} \neq 0.0$; specifically there is a far smaller chance of exhibiting non-monotonicity in the final training accuracy as the number of syndrome extractions increase, and a far smaller variance in the final training accuracy. Notably, when there is no noise on the ancilla qubits, the error detection works as expected and even at the highest noise levels, the training accuracy reaches $1.0$ within five rounds of syndrome extractions. As long as there is any noise on the ancilla qubits, the error detection loses effectiveness, and becomes unreliable at higher ancilla noise levels (i.e., $p_{anc}\geq0.005$). 

Thus, the high variability in final training accuracy at high ancilla error rates is very likely due to poor detection of ancilla errors. Since the ancilla qubits are entangled with the physical qubits, their errors spread easily into the physical qubits through the CNOT gates. While some of these errors will be detectable by syndrome measurements on the physical qubits, most of these errors will be undetectable. Hence, as the noise level increases, it becomes more difficult to protect the training through syndrome extractions, leading to the lower final accuracies and greater variance in its value. 

The plateauing in the mean final training accuracies observed at lower levels of noise exists as long as there is any ancilla noise present. At higher levels of ancilla noise, the plateauing is hidden by the greater variance and non-monotonicity in the evolution of the final training accuracy. The plateau indicates that there is a limit for how many ancilla-caused errors can be detected and removed from the physical qubits by the stabiliser code for a given Pauli Error Rate, leaving only errors that syndrome measurements cannot detect. When that limit is reached, adding more syndrome extractions will not result in the detection of more errors, leading to the plateau. 

\subsubsection{Threshold Ancilla Error Rate}
Our results for both noise models motivate the definition of a threshold Pauli Error Rate for ancilla qubits, such that when the error rate is larger than this threshold, error correction may not be effective. In particular, the addition of more syndrome extractions may not improve training, there is high variability in final training accuracy, and the maximum mean final training accuracy is considerably lower than $1.0$ for the system. 

In determining the threshold Pauli Error Rate for our system, we exclude ancilla error rates that produce a plateau at a final training accuracy of less than $0.90$, even if high variability is not an issue. We deduce from our results a threshold Pauli Error Rate of $p = 0.003$ for the gate noise model, and $p=0.004$ for the environmental depolarising noise model. For comparison, the current lowest single-qubit gate error rate exhibited by a NISQ device is $0.15\%$~\cite{Arute2019}, meaning that we may be able to run our simple QVC with the [[4,2,2]] code on the least noisy NISQ devices under special circumstances (for example, if the dominant noise is gate noise and environmental noise is very low). However, with the addition of real-world environmental noise to the gate noise, it is possible that the error threshold would be too low to run on currently available NISQ devices. 

There is a parallel between the threshold we have defined and the Threshold Theorem for quantum error correction, which asserts that there is a critical error rate below which sufficiently good quantum error correction codes can successfully correct errors. For error rates above this threshold, errors accumulate too quickly for effective error correction. However, despite the parallels, the Threshold Theorem does not explicitly cover the phenomenon of ancilla errors spreading to the physical qubits and reducing the effectiveness of the error correcting code.  

Though the threshold values we found are specific to our system and cannot be generalised to other systems, both the limit on the maximum training accuracy achievable and the existence of a threshold error rate for ancilla qubits should generalise to other combinations of QML algorithms and QEC codes. All QML algorithms contain rotation gates, and when implemented with QEC codes where ancilla qubits are needed for logically encoding rotation gates, thereby entangling the ancilla and physical qubits, we can expect error propagation between the ancilla and physical qubit registers. If the QEC code cannot correct the full range of errors that arise from such propagation, its effectiveness will be limited in noisy environments, especially above a threshold error rate. These results could also be relevant for implementing certain non-QML algorithms with QEC codes. This is because many QEC codes require ancilla qubits to logically encode rotation gates (and other non-Clifford gates), and since no known code can fully address the full spectrum of possible errors in practice, these implementations could also theoretically suffer from errors propagating from ancilla qubits to physical qubits. 

Our findings highlight the need for additional considerations when applying techniques to achieve fault-tolerant implementations of QML algorithms. Fault-tolerant quantum machine learning may only be achievable with QEC if additional error mitigation techniques are used with it. For example, performing rotations with error mitigation instead of as logical operations (such as with zero-noise extrapolation~\cite{Pascuzzi_2022} or dynamical gate error correction~\cite{PhysRevA.80.032314}), and using flag qubits normally applied to syndrome qubits~\cite{PhysRevX.10.011022, PRXQuantum.1.010302} in the ancilla register to minimise errors before they propagate, may help minimise uncorrectable errors. 

\section{Conclusion}
In this chapter, we discussed the application of quantum machine learning, specifically the variational circuit, in the context of a noisy environment, where extensive simulation of a whole training process is lacking in most literature. Established on the theory of quantum error correction, we proposed a partial QEC protocol by foregoing magic state distillation. In this way, we balanced the conflicting needs of minimal mitigation overhead and effectiveness against depolarizing noise. The simulation reveals that to achieve a reasonable logical error rate, the spacetime overhead to execute a simple 10-qubit variational circuit is at least two orders of magnitude more expensive compared to using the raw magic states directly. Remarkably, we simulated the QVC model performance, particularly its trainability, under the setting of partial-QEC with single-qubit gate noise only. In the experiments for classifying the handwritten numbers of MNIST datasets, we presented clear evidence of the model's trainability with a depolarizing noise strength of $p=1.47 \times 10^{-3}$, which corresponds to a gate error rate of $1.96 \times 10^{-3}$, a value higher than that of state-of-the-art quantum computers. Moreover, the poor performance when both single- and two-qubit noise was turned on further reinforces the necessity of partial quantum error correction instead of directly running on physical qubits.

Extending our investigation of QVC performance under noisy conditions, we directly incorporated quantum error detection into the training process and evaluated the resulting QVC accuracy. Specifically, we examined the performance of a simple parity-classifying 2-qubit QVC implemented with the [[4,2,2]] error-detecting code under Pauli noise. We demonstrated through classical simulations that even simple error detection codes can improve model performance. However, our findings suggest that the effectiveness of QEC is constrained by ancilla-induced error propagation, which sets a threshold on achievable QVC accuracy. 

Given these limitations and the resource demands of implementing QEC with QML, it is clear that a purely QEC-based approach to fault-tolerant QML is not feasible. To minimise resource usage and maximise QEC effectiveness, we require a hybrid approach combining QEC, algorithmic design and error mitigation strategies for the large-scale, practical implemention of QML on noisy quantum hardware.  

\begin{acknowledgement}
The authors greatly acknowledge the computational resources provided by the National Computing Infrastructure (NCI) through the National Computational Merit Allocation Scheme (NCMAS), the University of Melbourne's Research Computing Services, the Petascale Campus Initiative, and the Commonwealth Scientific and Industrial Research Organisation (CSIRO) High Performance Computing cluster. H.K. was supported by the Australian Government Research Training Program Scholarship, and E.A. was supported by the CSIRO Impossible Without You program. 
\end{acknowledgement}

\bibliographystyle{unsrt}
\bibliography{references}

\end{document}